\documentclass[11pt,a4paper]{article}
\usepackage{jcappub}
\usepackage{color}
\usepackage{graphicx}
\usepackage{times}
\usepackage{amssymb}
\usepackage{amsmath}
\usepackage{url}

\def\der{{\rm d}}

\def\msun{\textrm{M}_{\odot}}

\title{Probing the diffuse baryon distribution with the lensing-tSZ
 cross-correlation}

\author[a,b]{Yin-Zhe Ma,}
\author[b]{Ludovic Van Waerbeke,}
\author[b,c]{Gary Hinshaw,}
\author[b,d]{Alireza Hojjati,}
\author[b]{Douglas Scott,}
\author[a]{Joe Zuntz}

\affiliation[a]{Jodrell Bank Centre for Astrophysics, School of Physics and Astronomy, The University of Manchester, Oxford Road, Manchester, UK. M13 9PL}

\affiliation[b]{Department of Physics and Astronomy,
University of British Columbia, 6224 Agricultural Road, Vancouver, V6T 1Z1,
 BC Canada.} 

\affiliation[c]{Canada Research Chair in Observational Cosmology}

\affiliation[d]{Simon Fraser University, 8888 University Drive Burnaby,
 B.C. Canada V5A 1S6}

\emailAdd{mayinzhe@manchester.ac.uk}

\abstract{
Approximately half of the Universe's baryons are in a form that has been
hard to detect directly.  However, the missing component can be traced
through the cross-correlation of the thermal
Sunyaev-Zeldovich (tSZ) effect with weak gravitational lensing.
We build a model for this correlation and
use it to constrain the extended baryon component, employing data
from the Canada France Hawaii Lensing Survey and the
{\it Planck\/} satellite.  The measured
correlation function is consistent with an isothermal
$\beta$-model for the halo gas pressure profile, and the 1- and 2-halo
terms are both detected at the 4$\sigma$ level. In addition, we measure the
hydrostatic mass bias $(1-b)=0.79^{+0.07}_{-0.10}$,
which is consistent with numerical simulation results and the constraints
from X-ray observations.  The effective temperature of the
gas is found to be in the range ($7\times10^{5}$--$3 \times10^{8}$)\,K,
with approximately $50\%$ of the baryons
appearing to lie beyond the virial radius of the halos,
consistent with current expectations for the warm-hot intergalactic medium.}

\arxivnumber{1404.4808}


\begin{document}
\maketitle

\section{Introduction}

The general processes driving structure
formation, from the sizes of galaxies to the largest scales observable,
are reasonably well understood, though many details are still
unclear.  Knowledge of the distribution of baryonic and dark
matter in galaxies, groups, and clusters of galaxies is essential
for understanding how they form and evolve, including complex processes
such as down-sizing and star-formation
quenching \citep{benson2010,tinker2013}.  However, stellar mass accounts for
only ${\sim}\,10\%$ of all the baryons in the Universe; the other
$90\%$ resides in a diffuse component, a large fraction of which
is thought to reside in low mass halos~\cite{Fukugita04}.  A
complete picture of structure formation requires a full
census of baryons in the Universe.  Baryons are more dissipative than dark
matter, and hence naturally populate the centres of halos, but
feedback processes play a
fundamental role in recycling baryons back to a diffuse form.
Thus, accounting for the extended baryon distribution is necessary to
understand the physical processes governing
structure formation, including star formation and feedback.

Historically, the diffuse component is observed via its X-ray
emission or through the thermal Sunyaev-Zeldovich effect (tSZ,
i.e., inverse Compton scattering \cite{SZ1972}), but the
sensitivity of current instruments limits such observations to the
most massive, densest, and hottest gas environments.  To date,
only about half of the known baryon component has been directly
observed at redshifts less than $z \simeq 2$
\cite{Fukugita04,Bregman07}; the remaining baryons are thought to
be too cold to be detected with X-rays or the SZ effect, and too
warm to be detected in the UV.  Numerical simulations suggest that
the ``missing'' baryons might be in a warm, low-density plasma
(${\sim}\,10^{5}$--$10^{7}\,$K) correlated with large structures
and filaments \cite{Cen06}.

One possible way of observing these baryons is by
cross-correlating with another cosmic field. Gravitational lensing
by large-scale structure provides an unbiased tracer of the
matter distribution that can be used for this purpose.
\citet{Waerbeke14} found a significant
correlation between the Canada France Hawaii Lensing Survey
(CFHTLenS) mass map and tSZ maps
obtained from {\it Planck\/} satellite data. 
This signal was consistent with warm
baryonic gas tracing large-scale structure, with an amplitude
$\bar{n}_{\rm e} T_{\rm e} b_{\rm gas} \simeq 0.201 {\,\rm keV\, m^{-3}}$
at redshift $z=0$. This suggests that if the bias $b_{\rm gas}\simeq 6$
and $\bar{n}_{\rm e}=0.25 {\, \rm m^{-3}}$ (the
cosmic baryon abundance), then it is in line
with the missing baryons being at $T_{\rm e}\simeq10^{6}\,$K.

The model adopted for the warm gas in Ref.~\cite{Waerbeke14} was
simplistic and did not capture some of the essential physical
properties.  It assumed: (i) that the temperature and density of
the gas are independent of the underlying halo mass and redshift;
and (ii) that the bias of gas pressure relative to dark matter
follows $b_{\rm gas} \propto a$, where $a$ is the cosmic scale
factor, independent of halo mass and redshift.  Moreover the
formalism used could not account for gas lying outside single halos, thus
it was incapable of tracking the ``missing baryons'' that are
thought to reside outside the cluster virial radius.  Here we
attempt to provide a realistic description of the baryon
distribution within the framework of the ``halo
model''~\cite{Cooray02}. By interpreting the cross-correlation
between tSZ and lensing we investigate the consequences for the
warm baryonic component.

Except for Fig.~\ref{fig:wmap-planck}, we use best-fit cosmological
parameters obtained from the {\it Planck\/} satellite~\cite{Planck16}
throughout the paper, i.e., \{$\Omega_{\rm m}$, $\Omega_{\rm b}$,
$\Omega_{\Lambda}$, $\sigma_{8}$, $n_{\rm s}$, $h$\} =
\{0.3175, 0.0490, 0.6825, 0.834, 0.9624, 0.6711\}.

\section{Lensing-tSZ cross-correlation data}
The tSZ effect is produced by inverse-Compton scattering of cosmic microwave
background (CMB) photons off electrons in the hot intra-cluster
gas. At frequency $\nu$, this induces a temperature anisotropy
along the line-of-sight characterized by the Compton $y$-parameter,
\begin{eqnarray}
\frac{\Delta T}{T_0}=y \, S_{\rm SZ}(x), \text{ }y= \int n_{\rm
e}\sigma_{\rm T}\frac{k_{\rm B}T_{\rm e}}{m_{\rm e}c^{2}} {\rm d}l,
\end{eqnarray}
where $S_{\rm SZ}(x) = x\coth (x/2) -4$ gives the tSZ spectral
dependence \cite{SZ1972}.  Here, $x=h\nu/k_{\rm B} T_0$,
$n_{\rm e}$ is the electron density, $\sigma_{\rm T}$ is the
Thomson cross-section,  $T_{\rm e}$ is the electron temperature,
and $T_0$ is the present-day CMB temperature.

For this analysis, we use the cross-correlation data described in
Ref.~\cite{Waerbeke14}. The gravitational lensing mass map,
$\kappa$, is based on CFHTLenS data
\cite{Benjamin13,Heymans12,Hildebrandt12,Erben13,Miller13} and covers
$154\,{\rm deg}^2$ in four separate patches. The tSZ $y$ maps are
obtained from a linear combination of the four {\it Planck\/} channel maps
at 100, 143, 217, and 353\,GHz \cite{Planck1}. The $\kappa$ and
$y$ maps are smoothed
by a Gaussian beam with FWHM of $10$\,arcmin and $9.5$\,arcmin, respectively.
Several different $y$
maps were produced in order to test for contamination of the SZ
signal by thermal dust and CO line emission. These maps, labeled
B--H in Ref.~\cite{Waerbeke14}, were constructed using different
channel combinations, which would be expected to have very
different levels of signal contamination. As already noted, the
expected SZ contamination levels range over more than a factor of
6 across this set, while the measured cross-correlation signal
varies by less than $10\%$.  In this paper, we have adopted SZ map
D as our best estimate of the foreground-reduced $y$ map, because it projects
out dust assuming $\beta_{\rm dust}=1.8$, similar to the recent findings by
the {\it Planck\/} collaboration~\cite{Planck22}.  
However, as noted in Ref.~\cite{Waerbeke14} using different spectral indices for
the dust de-projection has only a 
10\% effect on our signal and does not affect our conclusions.

\section{$\kappa$-$y$ correlation function in the halo model}
The
lensing-tSZ cross-correlation power spectrum is the sum of two terms,
$C^{\kappa y}_{\ell} = C_{\ell}^{\kappa y,\textrm{1h}}
 + C_{\ell}^{\kappa y,\textrm{2h}}$, where ``1h'' and ``2h'' refer to
the 1- and 2-halo terms, respectively. The 1-halo term, the
Poissonian contribution, is given by~\cite{Cooray02}
\begin{eqnarray}
C_{\ell}^{\kappa y,\textrm{1h}}  =  \int^{z_{\rm max}}_{0}{\rm d}z
 \frac{{\rm d}V}{{\rm d}z{\rm d}\Omega}   \int^{M_{\rm max}}_{M_{\rm min}}{\rm d}M
 \frac{\der n}{\der M} y_{\ell}(M,z)\, \kappa_{\ell}(M,z),
\label{eq:1halo}
\end{eqnarray}
where $\der V/(\der z \der \Omega) = c \chi^{2}/H(z)$ is the
comoving volume per unit redshift and solid angle, with $\chi(z)$
the comoving distance to redshift $z$ (in the best-fit {\it Planck\/}
cosmology \cite{Planck16}).  The quantity $\der n/\der M$
is the halo mass function, taken here to be the
Sheth-Tormen form \cite{Sheth02}.  The multipole functions
$y_{\ell}$ and $\kappa_{\ell}$ are related to the halo gas and
mass profiles, respectively, as now described.

\begin{figure}
\centerline{\includegraphics[bb=150 0 550 300, width=2.3in]{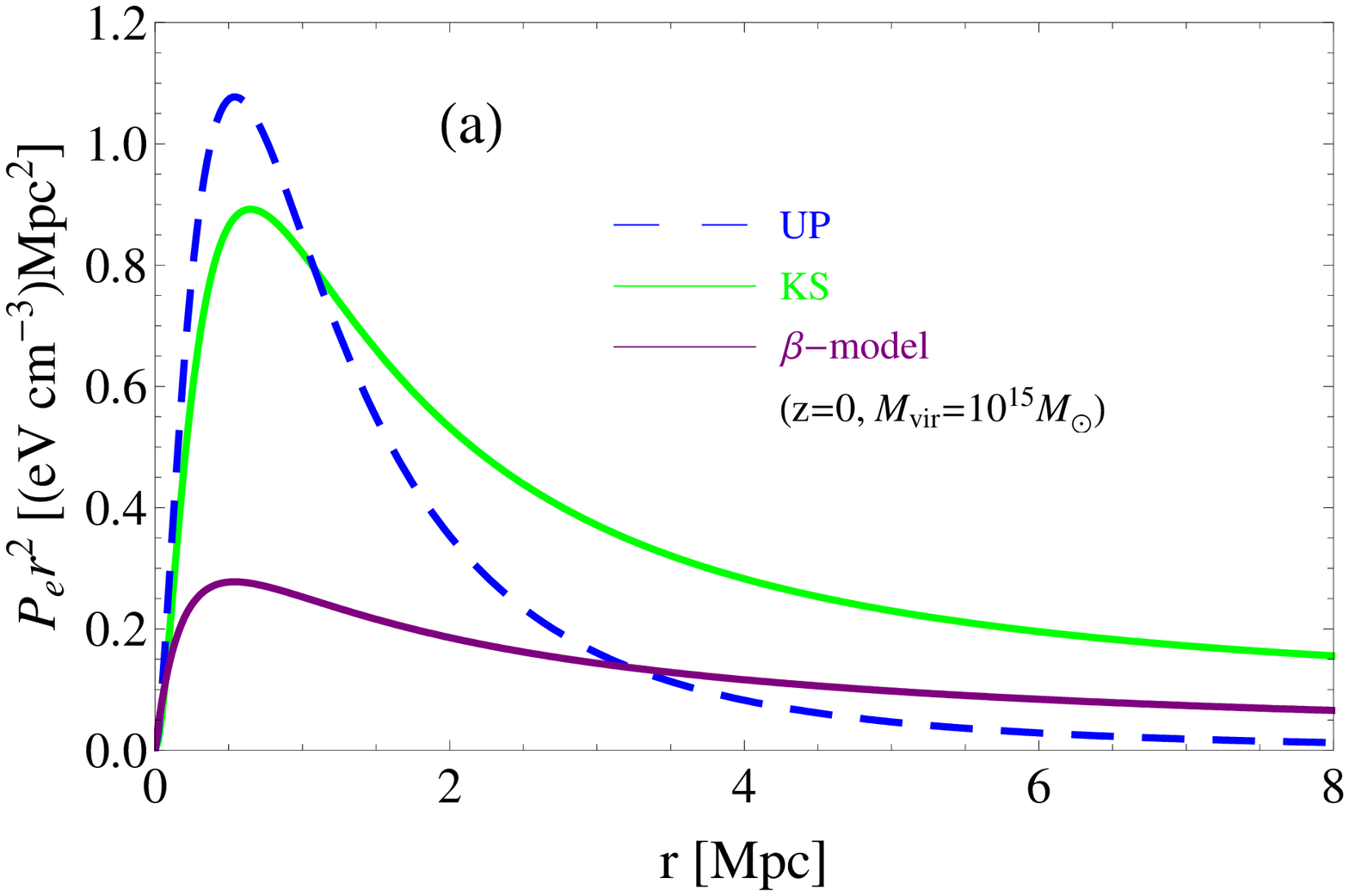}
\includegraphics[bb=20 0 500 300, width=2.4in]{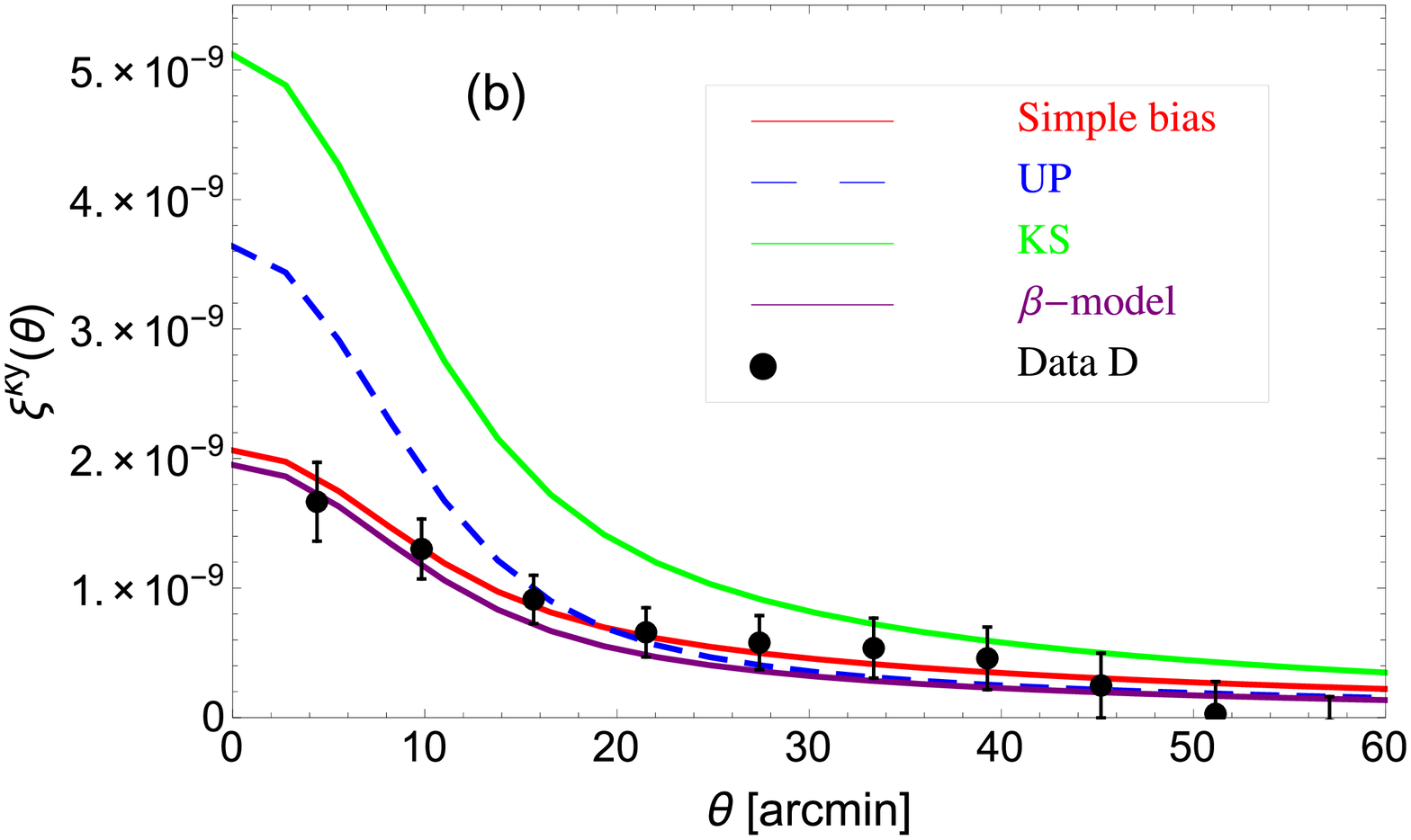}} \caption{Models for: (a) the
radial pressure profile within a halo; and (b) the $\kappa$--$y$
correlation function derived from the halo model. The pressure
profiles considered are: {\it red}, the simple bias model, with
$b_{\rm gas}T_{\rm e}n_{\rm e}=0.201\,{\rm keV}\,{\rm m}^{-3}$
\cite{Waerbeke14}; {\it blue dashed}, the universal pressure (UP)
profile \cite{Arnaud10}; {\it green}, the Komatsu-Seljak (KS)
profile \cite{Komatsu02}; and {\it purple}, the isothermal
$\beta$-model profile \cite{Waizmann09}. The data points in panel (b)
show the correlation function specifically for tSZ
data set ``D''.} \label{fig:model}
\end{figure}

\begin{figure}
\centerline{\includegraphics[width=3.3in]{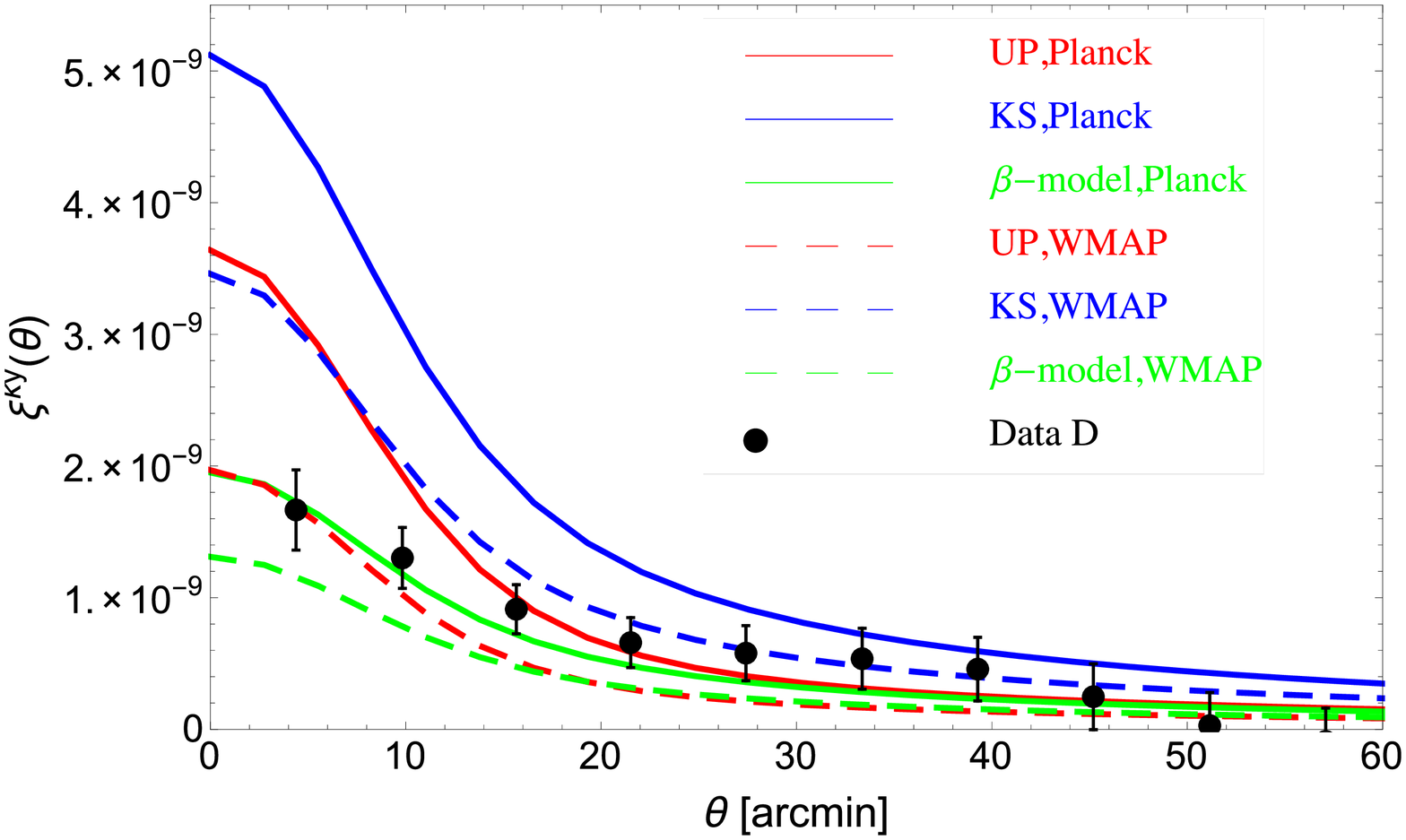}
\includegraphics[width=3.3in]{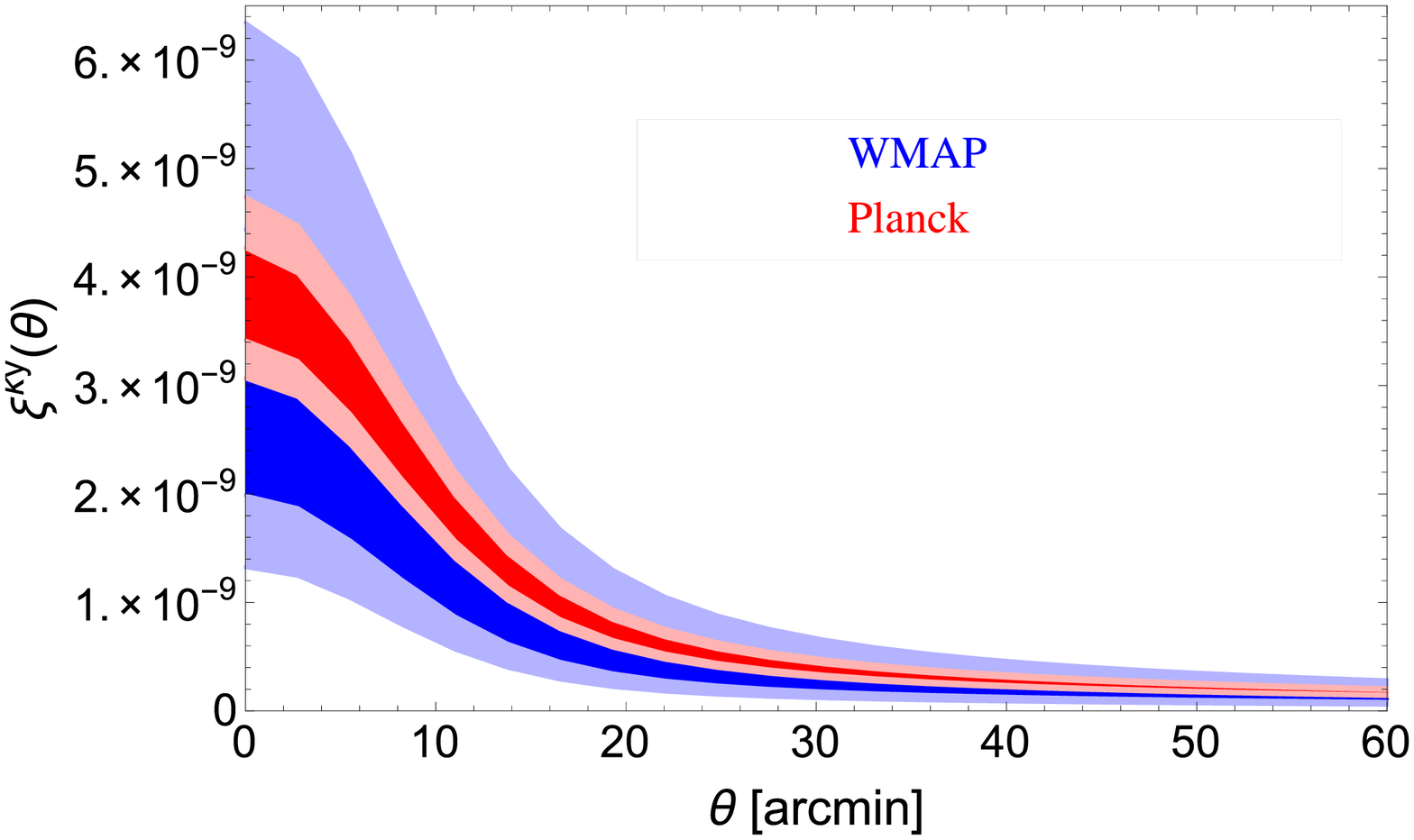}} 
\caption{{\it Left--}Comparison of the model predictions using best-fit
7-year {\it WMAP\/} and {\it Planck\/} 2013
cosmological parameters. The values are
\{$\Omega_{\rm m}$, $\Omega_{\rm b}$, $\Omega_{\Lambda}$, $\sigma_{8}$,
$n_{\rm s}$, $h$\} =
\{0.272, 0.0455, 0.728, 0.81, 0.967, 0.704\} for {\it WMAP\/} and
\{0.3175, 0.0490, 0.6825, 0.834, 0.9624, 0.6711\} for {\it Planck\/}.
{\it Right}-- the same comparison for the UP model, but sampling
the posterior distribution of cosmological parameter space. The deep (shallow)
blue and red regions are for {\it WMAP\/} and {\it Planck\/} 68\% (95\%)
confidence levels, respectively.} 
\label{fig:wmap-planck}
\end{figure}


The quantity $\kappa_{\ell}(M,z)$ is the Fourier transform of the convergence
profile of a single halo of mass $M$ and redshift $z$:
\begin{equation}
\kappa_{\ell}=\frac{W^{\kappa}(z)}{\chi^{2}(z)}
 \frac{1}{\bar{\rho}_{\rm m}}\int^{r_{\rm vir}}_{0}\der r (4 \pi r^{2})
 \frac{\sin(\ell r/\chi)}{\ell r/\chi} \rho(r;M,z). \label{eq:kappa-ell}
\end{equation}
Here $\bar{\rho}_{\rm m}$ is the comoving matter density,
$\rho(r;M,z)$ is the matter halo profile, taken here to be the
Navarro-Frenk-White (NFW) form~\cite{Navarro96}, and $W^{\kappa}$
is the lensing kernel, which is given in Eq.~(1) and plotted in
figure~1 of Ref.~\cite{Waerbeke14}. For the CFHTLenS data, the
kernel peaks at $z\simeq 0.37$; we have verified that Eq.~(\ref{eq:1halo}) has
converged at $z_{\rm max} = 3.0$ and adopt this redshift cutoff. For the integral over
mass we have verified that the integral has converged at a lower limit of
$10^{12}\msun$ and an upper limit of $10^{16}\msun$ and thus adopt this mass range.

The quantity $y_{\ell}(M,z)$ in Eq.~(\ref{eq:1halo}) is the 2-d Fourier
transform of the projected gas pressure profile of a single halo
of mass $M$ and redshift $z$  \cite{Planck21}:
\begin{equation}
 y_{\ell}=\frac{4\pi r_{\rm s}}{\ell^{2}_{\rm s}}
 \frac{\sigma_{\rm T}}{m_{\rm e}c^{2}} \int \der x \, x^{2}
 \frac{\sin(\ell x/\ell_{s})}{\ell x/\ell_{s}}
 P_{\rm e}(x;M,z). \label{eq:y-ell}
\end{equation}
Here $x=a(z)r/r_{\rm s}$, $\ell_{\rm s}=a\chi/r_{\rm s}$,
$r_{\rm s}$ is the scale radius of the 3-d pressure profile, and $P_{\rm e}$ is the electron pressure.
The ratio $r_{\rm vir}/r_{\rm s}$ is called concentration parameter,
which we take to be~\cite{Duffy08}
\begin{eqnarray}
c=\frac{5.72}{(1+z)^{0.71}}\left(\frac{M_{\rm vir}}{10^{14}h^{-1}\msun}
 \right)^{-0.081} \label{eq:value-c}.
\end{eqnarray}
This is based on the assumption that the ratio of $r_{\rm vir}/r_{\rm s}$
for the gas profile follows the same ratio as for the NFW (dark matter)
profile.  We have verified that the integral in Eq.~(\ref{eq:y-ell}) has
converged by $r=5r_{\rm vir}$ and adopt this as an upper cutoff radius.

The 2-halo term (to add to Eq.~\ref{eq:1halo}) is given by
\begin{eqnarray}
 C^{\kappa y,\textrm{2h}}_{\ell}  &= &  \int^{z_{\rm max}}_{0} \,
 \der z \frac{\der V}{\der z \der \Omega}\,P^{\rm lin}_{\rm m}(k=\ell/\chi,z) \\ \nonumber
& \times & \left[\int^{M_{\rm max}}_{M_{\rm min}}\der M
 \frac{\der n}{\der M}b(M,z) \kappa_{\ell}(M,z) \right]  \left[\int^{M_{\rm max}}_{M_{\rm min}}\der M
 \frac{\der n}{\der M}b(M,z) y_{\ell}(M,z) \right], \label{eq:2halo}
\end{eqnarray}
where $P^{\rm lin}_{\rm m}(k,z)$ is the 3-d linear matter power
spectrum at redshift $z$, which we obtained from the code {\sc
camb} \cite{camb}, with the best-fit parameters from {\it
Planck}~\cite{Planck16}.  Here $b(M,z)$ is the gravitational
clustering bias function (from Ref.~\cite{Mo02}).
In order to compare the halo model to the
cross-correlation data, $\xi^{\kappa y}(\theta)$, we Legendre
transform $C^{\kappa y}_{\ell}$ into real space.

\section{Pressure profile}
\subsection{Models for gas pressure}
We consider three different gas models in this study: the Komatsu-Seljak (KS)
profile \cite{Komatsu02}; the universal pressure (UP) profile
\cite{Arnaud10}; and the isothermal $\beta$-model
\cite{Arnaud09,Waizmann09,Hallman07,Plagge10}. For the KS model, we use
equations~(D4)--(D13) in~\cite{Komatsu11} to implement the pressure profile.
For the UP model, the pressure is given by
\begin{eqnarray}
P(x\equiv r/R_{500}) = 1.65 \times 10^{-3} E(z)^{\frac{8}{3}}
 \left(\frac{M_{500}}{3\times 10^{14}h_{70}^{-1}\msun}
 \right)^{\frac{2}{3}+\alpha_{\rm{p}}}
 \mathbb{P}(x) \, h_{70}^{2}\text{ }\left[\textrm{keV cm}^{-3}\right],
 \label{eq:unipres}
\end{eqnarray}
where $h_{70}=(h/0.7)$, $\alpha_{\rm{p}} =0.12$, and
$M_{500}=(4\pi/3)500\rho_{\rm c}(z)R^{3}_{500}$, i.e., the total mass within
the radius where the total density contrast is $500$. Here
$\mathbb{P}(x)$ is the generalized NFW model \citep{Arnaud10}
\begin{eqnarray}
\mathbb{P}(x) = \frac{P_0}{(c_{500} x)^{\gamma}
 \left[1+(c_{500} x)^{\alpha}\right]^{(\beta-\gamma)/\alpha}}, \label{px}
\end{eqnarray}
where $P_{0}$ is the overall
magnitude of the pressure profile, and $c_{500}, \gamma, \alpha$,
and $\beta$ determine the slope of the profile. We use the parameter set
$\{P_{0}, c_{500}, \alpha, \beta, \gamma\} = \{6.41, 1.81, 1.33, 4.13, 0.31\}$,
which is obtained as the best-fit values of 62 nearby massive
clusters~\cite{Planck-i5}.

For the isothermal $\beta$-model, 
\begin{eqnarray}
n_{\rm e}(r)=n_{\rm e0}\left[1+\left(\frac{r}{r_{\rm s}} \right)^{2}
 \right]^{-3\beta/ 2},
\end{eqnarray}
and we use $\beta=0.86$, which is consistent with the fits to the
X-ray surface brightness~\cite{Hallman07} and with fits to 15 stacked SZ
clusters from South Pole Telescope data~\cite{Plagge10}. The quantity
$r_{\rm s}=r_{\rm vir}/c$ is the scale radius that we use in
Eq.~(\ref{eq:value-c}), derived by assuming that the underlying dark matter
distribution follows the NFW profile.
For the central density, we fix the normalization with
$4\pi \int^{r_{\rm vir}}_{0} n_{\rm e}(r)r^{2}\der r =N_{\rm e}$,
where $N_{\rm e}=(1+f_{\rm H})M_{\rm vir}f_{\rm gas}/(2 m_{\rm p})$ ~\cite{Waizmann09}.
Here $f_{\rm H}=0.76$ is the hydrogen mass fraction, and
$f_{\rm gas}=\Omega_{\rm b}/\Omega_{\rm m}$ is the baryonic gas fraction of the
Universe. We use equation~(14) in Ref.~\cite{Waizmann09} as the temperature in the
$\beta$-model, where this equation was calibrated against 24 hydrodynamic cluster
simulations~\cite{Mathiesen01}. Fig.~\ref{fig:model}a
shows the pressure profiles for each model in a halo of mass
$M_{\rm vir} = 10^{15}\,\msun$ at $z=0$. Note that the amplitudes
of the KS and UP profiles are much higher than that of the $\beta$-model,
the latter being fairly consistent with the simple bias approach
from Ref.~\cite{Waerbeke14}.

\subsection{Hydrostatic mass bias}
\begin{figure}
\centerline{\includegraphics[width=3.1in]{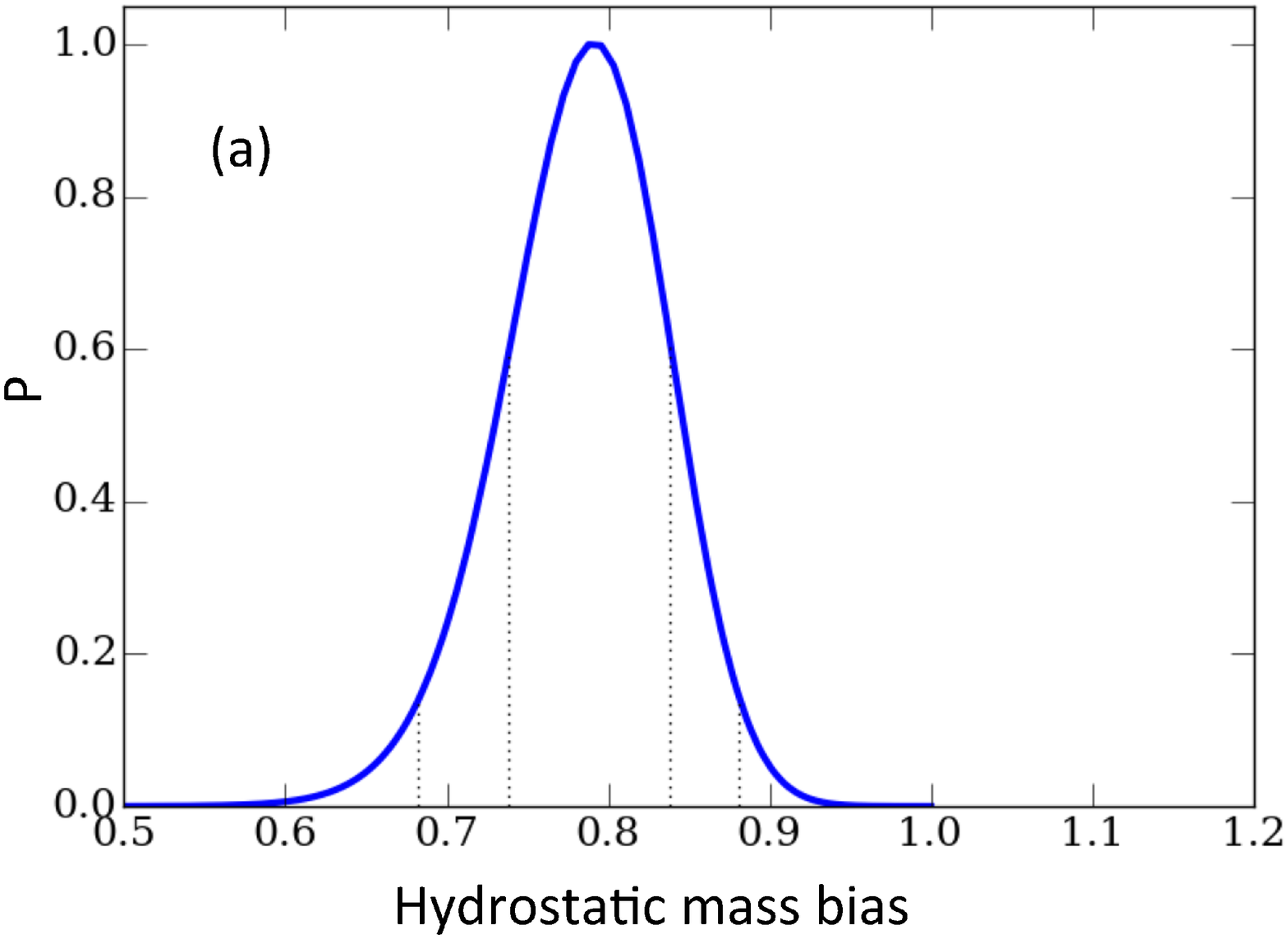}
\includegraphics[width=3.5in]{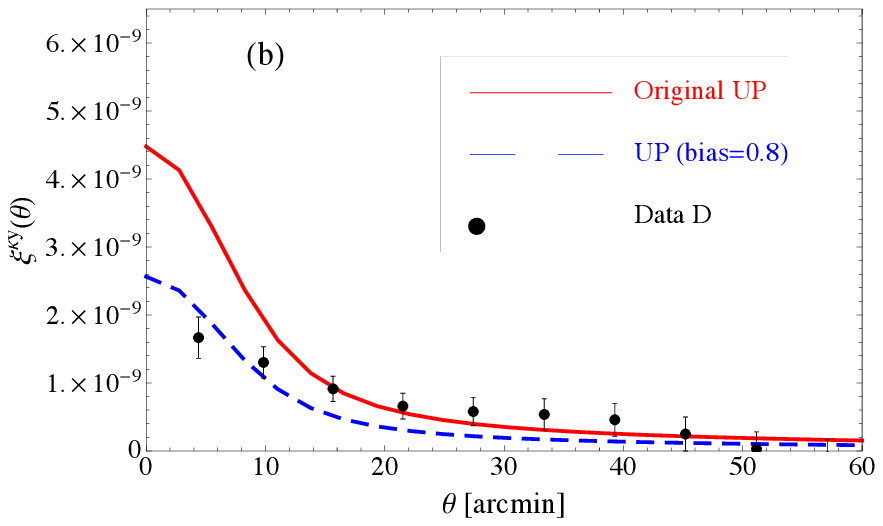}} 
\caption{{\it Left--}marginalized likelihood function for the hydrostatic
mass bias $1-b$ factor for data set ``D''  and the UP model. {\it Right}--
comparison between the original UP model and a UP model with $1-b=0.8$.} 
\label{fig:bias}
\end{figure}

\begin{table}
\begin{centering}
\begin{tabular}{@{}l c c c c }\hline
& Observation/Simulation& Quantity& Value ($68\%$ CL)& Reference\\ \hline 
& WtG& $1-b$& $0.688\pm0.072$& von der Linden et al.~2014~\cite{Linden14}\\
\cline{2-5}& CCCP& $1-b$& $0.76\pm0.11$& Hoekstra et al.~2015~\cite{Hoekstra15}\\
\cline{2-5}& {\it 400d} survey& $b$& $\simeq20\%$& Israel et al.~2014~\cite{Israel14}\\
\cline{2-5}& X-ray \& WL & $1-b$ & $0.66^{+0.07}_{-0.12}$ & Simet et al.~\cite{Simet15} \\
\cline{2-5} Data& CMB lensing& $1/(1-b)$& $0.99\pm0.19$& Planck 2015 results XXIV~\cite{Planck2015-24}\\
\cline{2-5} & tSZ--CMB Lensing & $1-b$ & $1.06^{+0.11}_{-0.14}$ & Hill \& Spergel~\cite{Hill14} \\
\cline{2-5}& CMB+SZ& $1-b$& $0.58\pm0.04$& Planck 2015 results XXIV~\cite{Planck2015-24}\\
\cline{2-5}& tSZ-Lensing correlation& $1-b$& $0.79^{+0.07}_{-0.10}$& This work\\ 
\cline{1-5}& Hydro-simulation& $b$& 10\%--20\%& Shaw et al.~2010~\cite{Shaw10}\\
\cline{2-5}& TreePM/SPH {\sc gadget}-3& $b$& $\sim 25\%$& Raisa et al.~2012~\cite{Rasia12}\\
\cline{2-5} Simu-& Eulerian cluster& $b$& 10\%--20\%& Nagai et al.~2007~\cite{Nagai07}\\
\cline{2-5} lations& N-body/SPH& $b$& 10\%--15\%& Piffaretti \& Valdarnini~2008~\cite{Piffaretti08}\\
\cline{2-5} & N-body/SPH& $b$& 5\%--20\%& Meneghetti et al.~2010~\cite{Meneghetti10}\\
\noalign{\vspace{-1.5pt}} \hline
\end{tabular}%
\caption{Comparison between measurements of hydrostatic bias from different
samples of real data and simulations. Here ``WtG'' is the Weighing the Giants
project~\cite{Linden14}, ``CCCP'' stands for Canadian Cluster
Comparison Project~\cite{Hoekstra15}, while ``400d'' stands for the
$400\,{\rm deg}^2$ Galaxy Cluster Survey Weak Lensing program, which claims
that their studies favour a small WL-X-ray mass bias, consistent with both
vanishing bias and $20\%$ bias~\cite{Israel14}. The ``tSZ--CMB Lensing'' row refers to the cross-correlation
between the thermal SZ map and the CMB Lensing map of {\it Planck}, and fitting the $(1-b)$ factor while fixing all other cosmological
parameters with {\it WMAP}9 values~\cite{Hill14}. The ``CMB+SZ'' case is not a
direct estimate of $1-b$, but gives the value required in order to reconcile
the tension between CMB and SZ determinations on the
$\sigma_{8}$--$\Omega_{\rm m}$ constraint.} \label{tab:b-value}
\end{centering}
\end{table}

Figure~\ref{fig:model}b (Fig.~\ref{fig:Cyy}a) shows the predicted
correlation functions (angular power spectrum), $\xi^{\kappa
y}(\theta)$ ($C^{\kappa y}_{\ell}$), for each of the pressure
profiles described above.  Also shown is the measured correlation
function using data set ``D'' from Ref.~\cite{Waerbeke14}.  It is
clear that the KS and UP models predict too much power at small
scales, while the isothermal $\beta$-model lies reasonably close
to the data at all scales.  Note that once the cosmological
parameters and gas model parameters have been chosen, a given gas model
amplitude and profile has no further freedom to be adjusted.

In Fig.~\ref{fig:bias}b, we plot the UP profile (as a red solid line), which predicts too much power on small angular scales compared to the data.
In comparison, we plot $M_{\rm obs,500}=(1-b)M_{\rm true,500}$ (as a dashed
blue line), where $1-b=0.8$ is the hydrostatic mass bias between observed and
true halo masses.
One can see that the
total amplitude is lowered, and the small angular scales fit better to the
data, while on larger angular scales the prediction is still lower than the
data. We run an MCMC chain to constrain this $1-b$ bias factor, and we
obtained the likelihood as shown in Fig.~\ref{fig:bias}a. The best-fit value is
$(1-b)=0.79^{+0.07}_{-0.1}\,$(at $95\%$ CL).

In Table~\ref{tab:b-value}, we make a comparison between measurements of
hydrostatic mass bias from real data and simulations. The simulations are listed
in the second half of the table. One can see that the simulations
consistently prefer a value of $b$ around $10$--$20\%$. For measurement from
real data, Ref.~\cite{Linden14} compares the {\it Planck\/} cluster mass with
the weak lensing mass from the WtG project for $22$ massive clusters, and find
that $(1-b)=0.688 \pm 0.072$, while Ref.~\cite{Hoekstra15} uses the same method
for $50$ clusters and finds a higher value. In addition, by simulating
{\it Planck} observations, Ref.~\cite{Melin15} proposes the method of comparing
CMB-measured mass to X-ray-measured mass, thus in Ref.~\cite{Planck2015-24}, by
using this method, the {\it Planck\/} collaboration finds the value
$1/(1-b)=0.99 \pm 0.19$, which lies towards the higher end of possible bias
values. Finally, in order to reconcile the apparent tension between the
cosmological parameter (in particular $\sigma_8$ and $\Omega_{\rm m}$)
between the CMB anisotropy measurements and SZ number counts,
Ref.~\cite{Planck2015-24} finds that the value of $(1-b)$ needs to be as low as
$0.58$; this is clearly much lower than the simulation results and individual
measurements.
 
Therefore, we conclude that our measurement of hydrostatic mass bias,
i.e., the $(1-b)$ value, is consistent with previous simulation results, and measurements from the CCCP and 400d surveys. However, the value we found is
slightly higher than the value found in the WtG project as well as the value necessary
to reconcile the tension between CMB and SZ cosmological constraints, and
slightly lower than the CMB lensing results; however, all are consistent
within $2\sigma$.

\subsection{{\it WMAP} and {\it Planck} cosmological parameters}
\label{sec:}
In the left panel of Fig.~\ref{fig:wmap-planck}, we plot the predictions of
$\xi^{\kappa y}(\theta)$ for three gas models by using the best-fit
cosmological parameters from the 7-year {\it Wilkinson Microwave Anisotropy
Probe} (i.e., {\it WMAP})~\cite{Komatsu11} and {\it Planck\/} 2013
results~\cite{Planck16}.
One can see that, since the {\it WMAP}-7 data prefer smaller values of
$\sigma_{8}$ and $\Omega_{\rm m}$ than {\it Planck}, the central value of the
correlation function drops somewhat for each different model. For the
KS, UP, and $\beta$ models, the central values of the correlation function drop
by about 33\%, 44\%, and 32\%, respectively. In the right panel, we sample the
entire cosmological parameter space by using the posterior sample chains
released for {\it WMAP\/} 9-year $\Lambda$CDM fits and the {\it Planck\/} 2015
``plik\,HM\,TTTEEE\,lowTEB'' chain, and plot the predicted uncertainties for
the UP model by using the software {\sc cosmosis}~\cite{Zuntz14}. The
68\% and 95\% band for $\xi^{\kappa y}$ are shown in deep and shallow colours,
respectively. One can see that, the CL bands of $\xi^{\kappa y}$ for
{\it WMAP\/} are much larger than for {\it Planck}, due to its larger parameter
uncertainties. However, since {\it Planck\/} results have smaller
uncertainties, in the following we will use the {\it Planck\/} cosmological
parameters. We therefore remind the reader that the subsequent
conclusions are based on the {\it Planck\/} cosmological parameters.

\section{Constraining the baryon component with the $\beta$-model}
\begin{figure}
\centerline{
\includegraphics[width=3.3in]{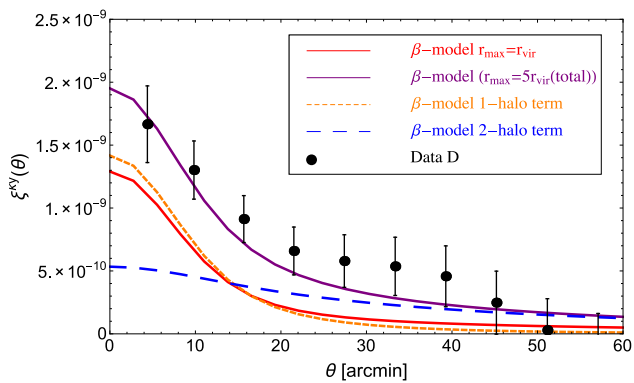}}
\caption{Separate contributions to the halo model using the
$\beta$-model pressure profile.  The dashed lines show the
contributions from the 1-halo ({\it orange}) and 2-halo ({\it
blue}) terms; the solid lines show the effect of truncating the
pressure profile integral in Eq.~(\ref{eq:y-ell}) at different
radii.} \label{fig:xi}
\end{figure}

\begin{figure}
\centerline{
\includegraphics[width=2.1in]{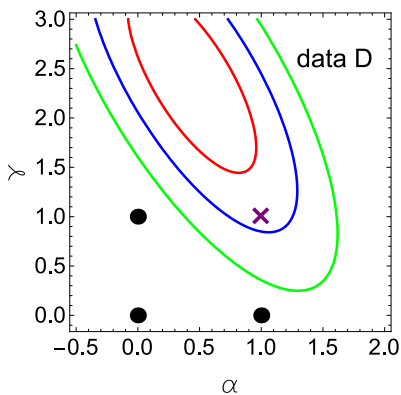}}
\caption{Joint constraints on $\alpha$ and $\gamma$ from
Eq.~(\ref{eq:chi2}) for specific data set D (other choices are
not dramatically different), showing the 68.3\%, 95.4\% and 99.7\%
confidence contours.  The model correlation function used here is
based on the isothermal $\beta$-model, and the nominal model
($\alpha=\gamma=1$) is indicated by the purple cross.  The black
dots are for models with no correlation ($\alpha=\gamma=0$) and
with separate 1-halo and 2-halo terms.} \label{fig:contourD}
\end{figure}

To study the $\beta$-model further, we separately examine the 1-halo 
and 2-halo contributions, as shown in Fig.~\ref{fig:xi}. 
The 2-halo term captures the effects of halo clustering, so it produces a
flatter contribution to $\xi^{\kappa y}(\theta)$ than the 1-halo term.  
To some extent, this term is a proxy for gas at large radii, 
not captured by the pressure profile in the 1-halo term.  
Any tendency for the data to favour a higher-than-predicted 2-halo
contribution might be pointing to the need for additional diffuse gas. As
expected, the 1-halo term dominates at small scales, while the 2-halo term
dominates at large scales, with a crossover point at $13$ arcmin. For an
average lens at $z\simeq 0.37$, this corresponds to a physical length of $4\,$Mpc.  
We quantify the relative contributions of the two terms by fitting each
with scaling coefficients $\alpha$ and $\gamma$:
\begin{eqnarray}
\chi^{2}(\alpha,\gamma)  =  \sum_{ij} \,
 \left[\xi^{\rm d}(\theta_{i})-\alpha \xi^{\rm 1h}(\theta_{i})-
 \gamma \xi^{\rm 2h}(\theta_{i})\right]  C^{-1}_{ij} \,
 \left[\xi^{\rm d}(\theta_{j})-\alpha \xi^{\rm 1h}(\theta_{j})
 -\gamma \xi^{\rm 2h}(\theta_{j})\right].
\label{eq:chi2}
\end{eqnarray}
Fig.~\ref{fig:contourD} shows the constraints on ($\alpha,\gamma$)
for the {\it Planck\/}-CFHTLenS cross-correlation using data set D
(although the other $y$-maps give similar results). Even though
the nominal model, $(\alpha,\gamma)\,{=}\,(1,1)$, is within the
95\% contour, the data prefer a fit with somewhat higher 2-halo
amplitude compared with 1-halo, which we interpret as an
indication that the $\kappa$--$y$ cross-correlation favours gas
that is further out in halos. Models with no correlation,
$(\alpha,\gamma)\,{=}\,(0,0)$, or with only 1-halo or 2-halo
contributions, $(\alpha,\gamma)\,{=}\,(1,0)$ or $(0,1)$,
respectively, are strongly rejected.  We quantify this for data
sets B--H in Table~\ref{tab1}, by evaluating $\Delta \chi^2
=\chi^2(\alpha,\gamma)-\chi^{2}_{\rm min}$, where $\chi^{2}_{\rm min}$
corresponds to the best-fit ($\alpha,\gamma$). We measure
the contributions from $1-$ and $2-$halo terms by calculating the
fractional area under the correlation function, i.e., $\int
\xi^{\kappa y,\rm 1h /2h}(\theta) \der \theta/\int \xi^{\kappa
y}(\theta)\der \theta$, finding that they each contribute about
50\% of the signal.

\begin{table}[tbp]
\begin{centering}
\begin{tabular}{cc@{\hskip 1em}c@{\hskip 1em}c}
\hline
\noalign{\vskip 3pt}
Data set & 2-halo only & 1-halo only & No correlation\\
\hline
\noalign{\vskip 3pt}
B& $4.8\times10^{-4}$& $6.7\times10^{-5}$& $4.5\times10^{-11}$\\
C& $1.9\times10^{-5}$& $2.0\times10^{-4}$& $1.3\times10^{-11}$\\
D& $1.1\times10^{-5}$& $1.5\times10^{-4}$& $1.5\times10^{-11}$\\
E& $2.4\times10^{-7}$& $7.9\times10^{-9}$& $1.0\times10^{-14}$\\
F& $4.7\times10^{-4}$& $2.7\times10^{-3}$& $1.0\times10^{-9\phantom{1}}$\\
G& $3.8\times10^{-3}$& $2.6\times10^{-2}$& $1.0\times10^{-7\phantom{1}}$\\
H& $8.3\times10^{-3}$& $1.3\times10^{-2}$& $2.4\times10^{-5\phantom{1}}$\\
\hline
\end{tabular}
\caption{For each $y$-map B--H, we list the probability that the fit in
Eq.~(\ref{eq:chi2}) allows: $\alpha=0,\gamma=1$ (no 1-halo term,
column 2); $\alpha=1,\gamma=0$ (no 2-halo term, column 3); and
$\alpha=\gamma=0$ (no cross-correlation, column 4).   We assume
$P=\exp(-\Delta\chi^{2}/2)$.} \label{tab1}
\end{centering}
\end{table}

To further probe contributions from gas at large radii and
in low-mass halos, we segregate the integrals in the $\beta$-model 
by mass and radius.  First, we truncate the gas distribution at one virial
radius in Eq.~(\ref{eq:y-ell}): $x_{\rm max} = a(z) r_{\rm vir}/r_{\rm s}$, 
as shown in Fig.~\ref{fig:xi}.  Next, we divide the model contributions into
two mass bins, 
$10^{12}$–-$10^{14}\msun$ and $10^{14}$-–$10^{16}\msun$, 
and two radial bins, $r \leq r_{\rm vir}$ and $r \geq r_{\rm vir}$. 
The fractional contributions to the integrated signal are presented in
Table~\ref{tab:percent}.  
Given our model assumptions, nearly half of the integrated signal (46\%) 
originates from baryons outside the virial radius of dark matter halos, while
40\% originates from low-mass halos.  One can additionally calculate the
fraction of baryons found inside the virial radius, $f = \int^{r_{\rm vir}}_{0} n_{\rm e}(r)r^{2}\der r/ \int^{\infty}_{0}
 n_{\rm e}(r)r^{2} \der r$.  
This $f$ function is redshift and mass dependent, but if we take the median
value of the redshift distribution $z=0.37$ and median mass $10^{14}\msun$ into
the above expression, we find $f = 35\%$, meaning that 65\% 
of the baryons are, on average, located beyond the virial radius.

One point to recall about our $\beta$-model predictions is that we assume the
gas to be isothermal, with temperature calibrated against 24 hydrodynamic
cluster simulations~\cite{Mathiesen01}.  Although this assumption is less
likely to hold for gas outside the virial radius, 
Ref.~\cite{Mathiesen01} shows that this single temperature model produces an
excellent fit to photon spectra.If the outer gas is cooler than the inner gas,
we would have to boost the gas 
density to retain the $\kappa$--$y$ signal we observe, and vice-versa.  
Note that, for halos at redshift $z=0.37$ (the mean probed by the CFHTLens
sample), the effective temperature ranges between
$T_{\rm vir} \simeq 7\times 10^{5}$ K and $3 \times 10^{8}$ K 
for halos in the range $10^{12}$--$10^{16} \msun$
(equation (14) in~\cite{Waizmann09}).  
The lower end of this range agrees with the expected temperature of 
the warm phase of the intergalactic medium residing in filaments and sheets of
clustered matter.  Thus it is plausible that our measurement is, in fact,
probing warm gas associated with lower mass halos that could constitute the
missing baryons.

\begin{table}[tbp]
\begin{centering}
\begin{tabular}{c|c@{\hskip 1em}c}
\rule{0pt}{10pt}
& $10^{12}\,\msun$--$10^{14}\,\msun$& $10^{14}\,\msun$--$10^{16}\,\msun$\\
\hline
\rule{0pt}{10pt}
(0.01--1)\,$r_{\rm vir}$& 26\%& 28\%\\
(1--100)\,$r_{\rm vir}$& 14\%& 32\%\\
\end{tabular}
\caption{Fractional contribution to the model cross-correlation function
arising from different mass and radial profile cuts.}
\label{tab:percent}
\end{centering}
\end{table}

\begin{figure*}
\centerline{
\includegraphics[bb=0 0 554 348, width=3.4in]{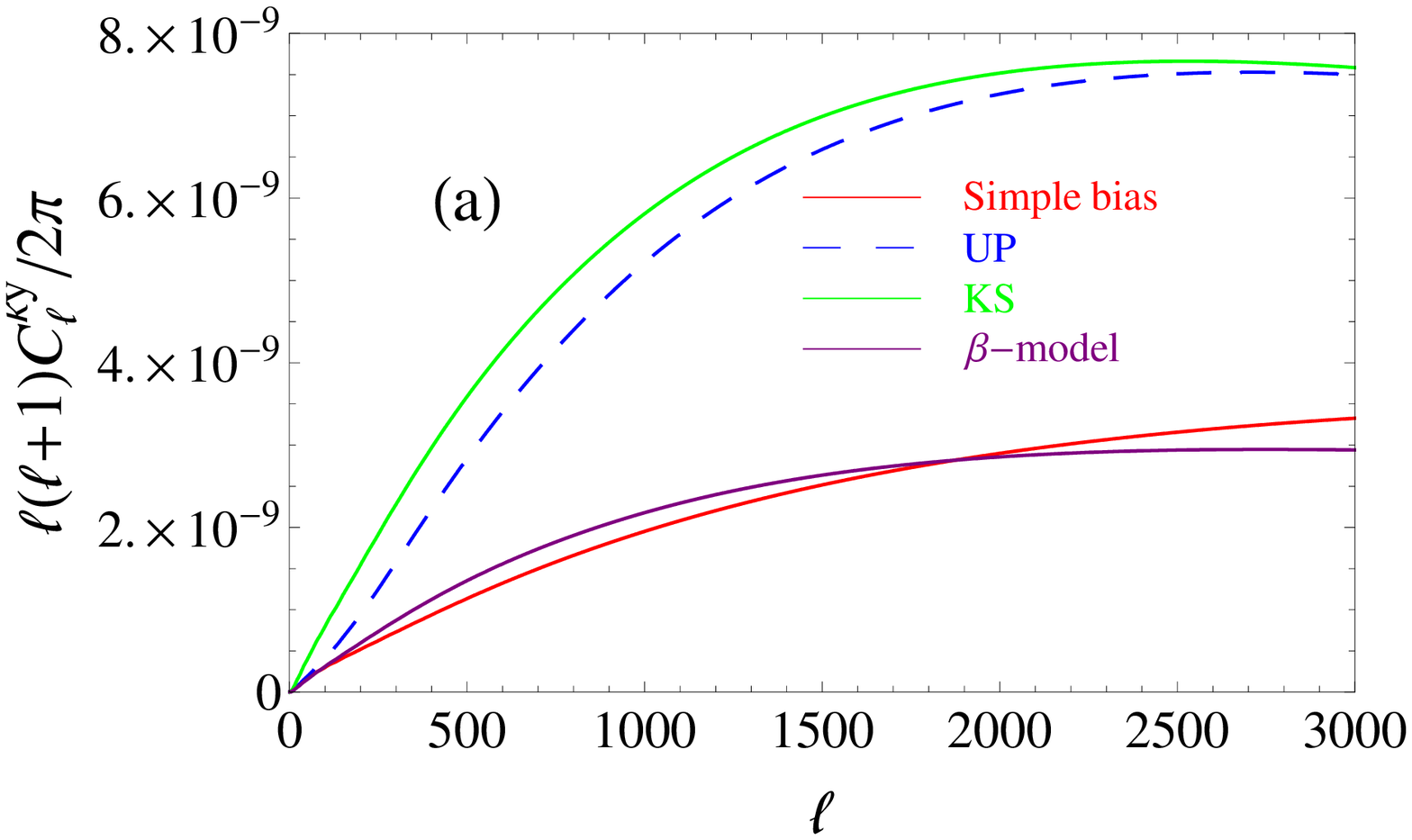}
\includegraphics[bb=7 10 490 291, width=3.4in]{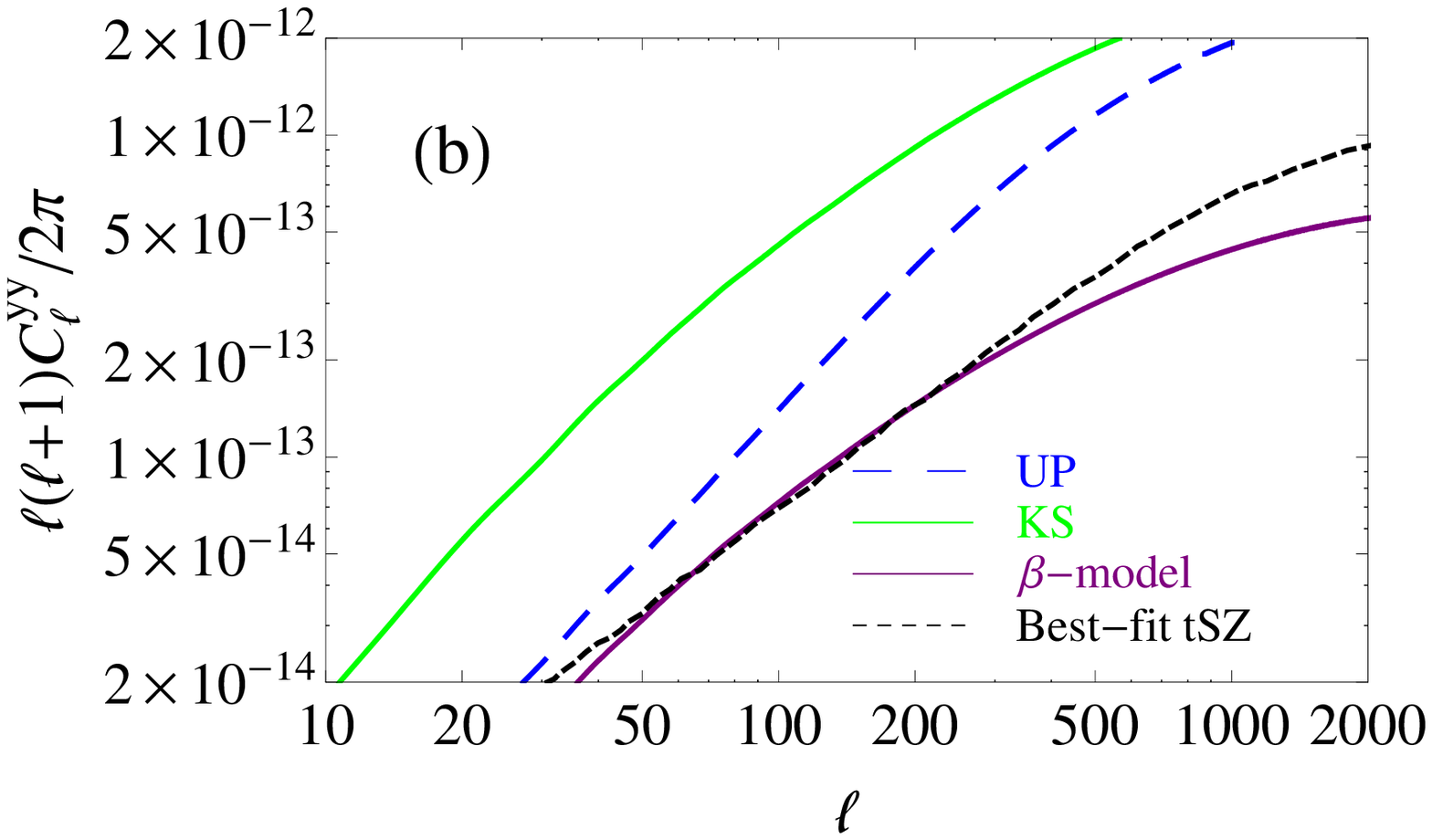}}
\caption{Predictions of the halo model for (a) $C_{\ell}^{\kappa y}$
and (b) $C_{\ell}^{yy}$.  The prediction for the
auto-correlation, $C_{\ell}^{yy}$, using the $\beta$-model
profile, agrees reasonably well with the best-fit tSZ spectrum
measured by {\it Planck}~\cite{Planck21} (dashed line).}
\label{fig:Cyy}
\end{figure*}

Our analysis of the cross-correlation signal can also be used to
predict the tSZ power spectrum $C_{\ell}^{yy}$, which can then be
compared to the measurement made by the {\it Planck\/} team
\cite{Planck21}.  Replacing $\kappa_{\ell}$ by $y_{\ell}$ in
Eqs.~(\ref{eq:1halo}) and (\ref{eq:2halo}), and using the
$\beta$-model for the pressure profile, we show our predicted
$C_{\ell}^{yy}$ in Fig.~\ref{fig:Cyy}b. The agreement with the
power spectrum derived directly from the {\it Planck\/} maps is
quite good, while the predictions based on the KS and UP profiles
are clearly too high. Note that our prediction is only correct if
the correlation coefficient $r$ between the 3-d pressure
and matter distributions is $1$. Hydrodynamical simulations
in Ref.~\cite{SBP2001} find $r\,{\sim}\,0.5$, but this conclusion is
still uncertain, so our prediction should only be regarded as a
lower limit.

\section{Discussion and Conclusions}
Our halo model for the
lensing--tSZ cross-correlation signal $\xi^{\kappa y}$ has enabled
us to investigate the baryon distribution at cluster scales and to
explore the possible identification of the missing baryons in the
warm-hot intergalactic medium (WHIM). The observed
cross-correlation function from the CFHTLenS mass map and the
{\it Planck\/} tSZ map is particularly effective at tracing baryons over a
wide range of clustering scales.

In the context of the universal pressure profile, we find that their predicted
$\xi^{\kappa-y}(\theta)$ function is higher than the observational data at
small angular scales; the added hydrostatic mass bias $(1-b)\simeq 0.8$ can
reconcile the tension to some extent, but on large angular scales it predicts
lower power than seen observationally. By employing a likelihood function to
fit the $(1-b)$, we find its value to be $(1-b)=0.79^{+0.07}_{-0.1}$
(at $95\%$ CL), which is consistent with previous values found values in
numerical simulations~\cite{Shaw10,Rasia12,Nagai07,Piffaretti08,Meneghetti10},
as well as some observational
constraints~\cite{Hoekstra15,Israel14,Planck2015-24}.

In the context of the isothermal $\beta$ profile, the 1- and 2-halo terms are each
detected at ${\sim}\,4\sigma$, while the total signal is
detected at ${\sim}\,6\sigma$.  We find evidence that baryons are
distributed beyond the virial radius, with a temperature in the
range of ($10^{5}$--$10^{7}$)\,K, consistent with the hypothesis
that this signal arises from the missing baryons.
We further separate the model signal into
different radial profile and mass bins, and find that about half of the
integrated signal arises from gas outside the virial radius of the
dark matter halos, and that 40\% arises from low-mass halos.

Our study is an example of a general class of large-scale
cross-correlations that are now becoming feasible, thanks to the
availability of deep multi-waveband surveys over large fractions
of the sky. Correlation of tSZ maps with galaxies
\cite{PlanckI11}, with CMB lensing \cite{Hill14} and with X-rays
\cite{Hajian13}, plus the use of correlations with the kinetic SZ
effect \cite{Hand12,Li2014}, allow for a multi-faceted study of
the role of baryon physics in structure formation. Further
improvements in the quality of the data will require more
sophisticated models than we have presented here, perhaps
involving direct comparison of diagnostics of the WHIM with
hydrodynamical simulations. Our results show that such
cross-correlation studies have the potential to trace the ``missing
baryons'' and to account for the cosmic baryon distribution with high precision.

\acknowledgments
We would like to thank Anna Bonaldi, J.~Colin Hill, Houjun Mo,
Pengjie Zhang,
and Eiichiro Komatsu for helpful discussions.  This paper made use of
the Planck Legacy Archive (\url{http://archives.esac.esa.int/pla}) and the
Canada-France-Hawaii Telescope Lensing Survey
(\url{http://www.cfhtlens.org/}).

\end{document}